\documentclass[12pt,twoside,a4paper]{article}
\usepackage[english]{babel}
\usepackage[T1]{fontenc}
\usepackage[utf8]{inputenc}
\usepackage{pgf}
\usepackage[sorting=none, sortcites=true,doi=false,url=false,    giveninits=true,maxbibnames=5, style=numeric-comp]{biblatex}

\usepackage{csquotes}
\bibliography{references}
\usepackage{amsmath,amssymb,amsfonts}
\usepackage{xcolor}
\usepackage{graphicx}
\usepackage[caption=false, label font=bf, labelformat=simple]{subfig}

\usepackage{standalone}
\usepackage{placeins}
\usepackage{siunitx}
\usepackage{suffix}
\DeclareSIUnit{\million}{\text{Mio.}}
\DeclareSIUnit\Molar{\textsc{m}}
\DeclareSIUnit\Gray{\textsc{Gy}}
\DeclareSIUnit\angstrom{\text {Å}}
\usepackage[version=4]{mhchem}

\DeclareUnicodeCharacter{0302}{\^}
\DeclareUnicodeCharacter{2212}{-}
\usepackage{authblk}
\usepackage[font=small,labelfont=bf,labelsep=period]{caption}

\long\def\RC#1\par{\makebox[0pt][r]{\bf RC:\hspace{4mm}}\textbf{\textit{#1}}\par} 
\WithSuffix\long\def\RC*#1\par{\textbf{\textit{#1}}\par} 
\long\def\AR#1\par{\makebox[0pt][r]{AR:\hspace{10pt}}#1\par} 
\WithSuffix\long\def\AR*#1\par{#1\par} 

\begin{document}
\title{Iron Oxide Nanoparticles as a Contrast Agent for Synchrotron Imaging of Sperm}
\author[1,2,*]{Mette Bjerg Lindh\o j}
\author[3]{Susan Rudd Cooper}
\author[3]{Andy S. Anker}
\author[4]{Anne Bonnin}
\author[5]{Mie Kristensen}
\author[6]{Klaus Qvortrup}
\author[7]{Kristian Almstrup}
\author[3]{Kirsten M. \O.\ Jensen}
\author[2,8]{Tim B. Dyrby}
\author[1]{Jon Sporring}
\affil[1]{Department of Computer Science, University of Copenhagen, Copenhagen, Denmark.}
\affil[2]{Danish Research Centre for Magnetic Resonance, Center for Functional and Diagnostic Imaging and Research, Copenhagen University Hospital Hvidovre, Hvidovre, Denmark.}
\affil[3]{Department of Chemistry and Nano-Science Center, University of Copenhagen, Copenhagen, Denmark.}
\affil[4]{Paul Scherrer Institut, Villigen PSI, Switzerland.}
\affil[5]{Department of Pharmacy, University of Copenhagen, Copenhagen, Denmark.}
\affil[6]{Department of Biomedical Science/CFIM, University of Copenhagen, Copenhagen, Denmark.}
\affil[7]{Department of Growth and Reproduction, Copenhagen University Hospital - Rigshospitalet, Copenhagen, Denmark.
}
\affil[8]{Department of Applied Mathematics and Computer Science
Technical University of Denmark, Kongens Lyngby
Denmark.}
\affil[*]{memo@di.ku.dk}
\date{}   
\maketitle
\begin{abstract}
Fast phase-contrast imaging offered by modern synchrotron facilities opens the possibility of imaging dynamic processes of biological material such as cells. Cells are mainly composed of carbon and hydrogen, which have low X-ray attenuation, making cell studies with X-ray tomography challenging. At specific low energies, cells provide contrast, but cryo-conditions are required to protect the sample from radiation damage. Thus, non-toxic labelling methods are needed to prepare living cells for X-ray tomography at higher energies. We propose using iron oxide nanoparticles due to their proven compatibility in other biomedical applications. We show how to synthesize and attach iron oxide nanoparticles and demonstrate that cell-penetrating peptides facilitate iron oxide nanoparticle uptake into sperm cells. We show results from the TOMCAT Nanoscope (Swiss Light Source), showing that iron oxide nanoparticles allow the heads and midpiece of fixed sperm samples to be reconstructed from X-ray projections taken at \SI{10}{\kilo\eV}.
\end{abstract}
\emergencystretch=1em


\section{Introduction}
Synchrotron X-ray tomography (SXRT) is a vital tool in scientific research of biological samples offering insight into diverse domains such as axon and muscle fibre morphology~\cite{Andersson2020, Borg2019} and the biomechanics of insects~\cite{Mokso2015, DosSantosRolo2014, Walker2014}. Combining the spatial resolution to view axons with the temporal resolution needed to see the dynamics of a blowfly offers excellent possibilities for studying the biomechanics of single cells.\par
Achieving dynamic tomography of single cells is dependent on two essential factors: High spatio-temporal resolution and high sample contrast at a low absorbed dose.\par
The recent development of brilliant synchrotron light sources and fast detector systems have made it possible to capture much faster motion and a wider range of sizes than ever seen before~\cite{Mokso2017}. Within the last decade, the effective pixel size has been pushed to the micro-and nanometer scale with sub-second exposure times~\cite{Mokso2017, Mokso2015, Kazantsev2017, Gonzalez-Tendero2017, Buurlage2019, Kuan2020, Zefreh2016, Rau2017, Villanova2017, Flenner2020}.\par
Mammalian sperm are very suitable candidates for probing the limits of SXRT for dynamic single-cell imaging as these are primary mammalian cells readily available and among the most dynamic cells known. Furthermore, changes in sperm motility caused by environmental chemicals acting on CatSper, the primary calcium channel of sperm, has been proposed to be a potential mechanism involved in recent decreasing fertility trends~\cite{Schiffer2014, Rehfeld2018}. Dynamic single-cell studies of sperm is highly relevant for understanding parameters important for fertility.
\par
The image intensity value in an X-ray projection comes mainly from a reduction of the intensity of the X-ray beam as it passes through the material, known as attenuation. The attenuation is caused by absorption or scattering of the beam and is affected by factors such as the atomic properties of the material and the beam energy. At most energies, cells and water-like substances, such as cell media needed for cells to survive, have very similar attenuation properties. This makes SXRT of living cells challenging as they provide almost no X-ray contrast against their surroundings. The water window is a unique spectral region where the attenuation difference between carbon and oxygen is especially large~\cite{Carzaniga2014}, allowing cells to be clearly distinguished from a water background~\cite{Jacobsen1998}. However, the increased absorption also increases the radiation dose deposited in the cells. Therefore, vitrification of the sample is required to protect the sample from radiation damage during X-ray tomography in the water window, making the study of living cell dynamics impossible.\par 
Different materials can be added to cells to achieve sufficient contrast in X-ray imaging at higher energies where the absorbed dose is less. One such method is the addition of electron-dense osmium tetroxide to myelin for enhanced contrast of axons~\cite{Andersson2020}. While osmium also provides good contrast in sperm, it is highly toxic, making it unfit for living cells. Other known contrast agents for X-ray imaging are iodine or gadolinium-based drugs, used for contrast enhancement during, e.g., angiography~\cite{GadIod, Wichmann2013} in medical CT scans. Gold nanoparticles can also provide contrast enhancement during X-ray imaging  ~\cite{nanoparticles, GoldNano, Cormode2011}.  Unfortunately, gold nanoparticles have adverse effects on the motility of sperm, leaving approximately one-quarter of the cells immotile after treatment~\cite{goldparticles}.\par
Iron oxide nanoparticles are widely used in biomedicine due to their high biocompatibility and magnetic properties. Furthermore, they enter sperm through endocytosis without affecting their motility or ability to fertilize an egg~\cite{Makhluf2006}.
Iron oxide attenuates the beam more than cells, making it a promising candidate for enhanced contrast in SXRT of biological samples such as sperm.\par 

When using nanoparticles for contrast enhancement during SXRT, the number of particles taken up by the cell, and their ability to distribute throughout the cell, are essential to provide sufficient contrast. To aid the uptake of the nanoparticles, cell-penetrating peptides (CPP), such as Penetratin, can be used~\cite{helix}. CPPs comprise a family of short peptides (up to approximately 30 amino acids), which inherently permeates cell membranes. Several CPPs, including Penetratin, are classified as cationic/amphipathic~\cite{cpp}. They internalize into cells via direct membrane translocation or endocytosis and have demonstrated the ability to facilitate uptake of co-incubated cargo such as nanoparticles into various cell types~\cite{tailwagging}. Penetratin has earlier demonstrated efficient internalization into sperm without compromising their cell viability or motility~\cite{Jones}.\par

\emph{We performed experiments to investigate whether iron oxide nanoparticles could be used as a non-toxic alternative to existing staining methods for achieving the necessary spatial resolution to image sperm using SXRT. To develop the staining method, we synthesized iron oxide nanoparticles covered in polyvinyl alcohol and incubated them with boar sperm, using Penetratin to facilitate the uptake. We used transmission electron microscopy to verify the attachment of the nanoparticles to the sperm and obtained SXRT volumes of fixed sperm stained with iron oxide nanoparticles at the TOMCAT beamline (
Swiss Light Source (SLS)) to assess the success of the staining.}



\section{Results}

\subsection{The synthesized nanoparticles consisted of a magnetite core coated in PVA}
\begin{figure}
\centering
\noindent\includegraphics[width=\linewidth]{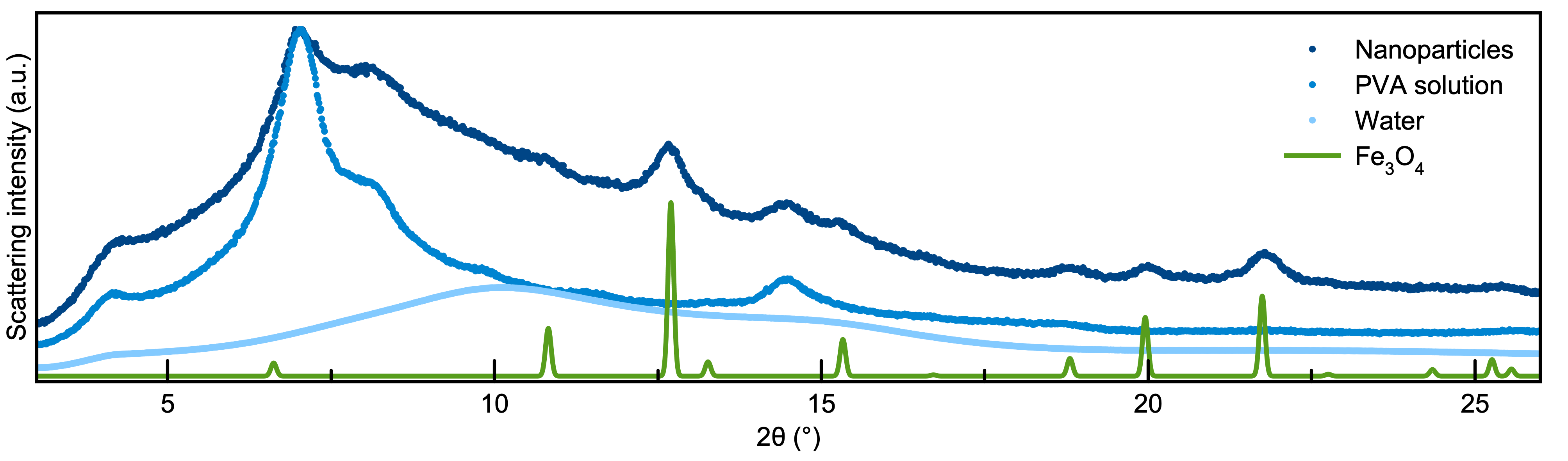}
\caption{\textbf{PXRD analysis.}  The graph shows a comparison of PXRD data of the solid nanoparticles before being thoroughly dried (dark blue), a PVA solution (medium blue), water (light blue), and a calculated scattering pattern of a published magnetite structure (green)~\cite{fleet}. The data are collected by measuring the scattering intensity at different scattering angles ($\mathrm{2\theta}$). The peaks from the scattering pattern of our synthesized nanoparticles (dark blue) coincide with the peaks from the PVA and iron oxide scattering patterns (medium blue and green) and are not present in the water background, indicating that the particles are iron oxide nanoparticles covered in PVA.}
\label{fig:pxrd}
\end{figure}
To make the non-toxic iron oxide nanoparticle contrast agent for boar sperm, we synthesized iron oxide nanoparticles coated in polyvinyl alcohol (PVA) by modifying a previously reported co-precipitation method~\cite{babes}, see Materials and Methods. Then, to verify the method's success, we used Powder X-ray diffraction (PXRD) to estimate the size of the particles and verify their crystal structure. Powder X-ray Diffraction (PXRD) is used to identify and characterize the structure of crystalline materials. The PXRD peaks contain information on the atomic arrangement in materials.\par
Figure~\ref{fig:pxrd} shows the X-ray diffraction pattern of the nanoparticles compared to the calculated X-ray scattering patterns of magnetite~\cite{fleet}. The X-ray scattering pattern from the nanoparticle suspension shows Bragg reflections (peaks) at values of 2$\theta$ equal to $12.7^\circ$, $15.3^\circ$, $18.8^\circ$, $20.0^\circ$ and $21.8^\circ$, which are also in the calculated magnetite pattern. In addition, in the X-ray scattering pattern from the nanoparticle suspension, we see large, broad peaks centred at $2\theta$ equal to $7.0^\circ$, $8.0^\circ$ and a peak at $2\theta$ equal to $14.5^\circ$, which are also present in the X-ray scattering pattern from the PVA solution, but not in the X-ray scattering pattern from the water background. This suggests that PVA is in the particle suspension, and we expect it to cover the surface of the nanoparticles. Using the Scherrer equation, equation~(\ref{eq:YY}) in Materials and Methods, we estimated the crystallite size of the particles to be approximately \SI{6}{\nano\metre} in diameter. Conclusively, PXRD analysis confirms that iron oxide nanoparticles with a size of approx. 6 nm have been synthesized, and that they are most likely covered by PVA. Combined with the small size of the particles, they seem to be a promising candidate for an X-ray contrast agent.

\subsection{Uptake of nanoparticles by sperm}
\begin{figure}
\centering
\noindent\includegraphics[width=\linewidth]{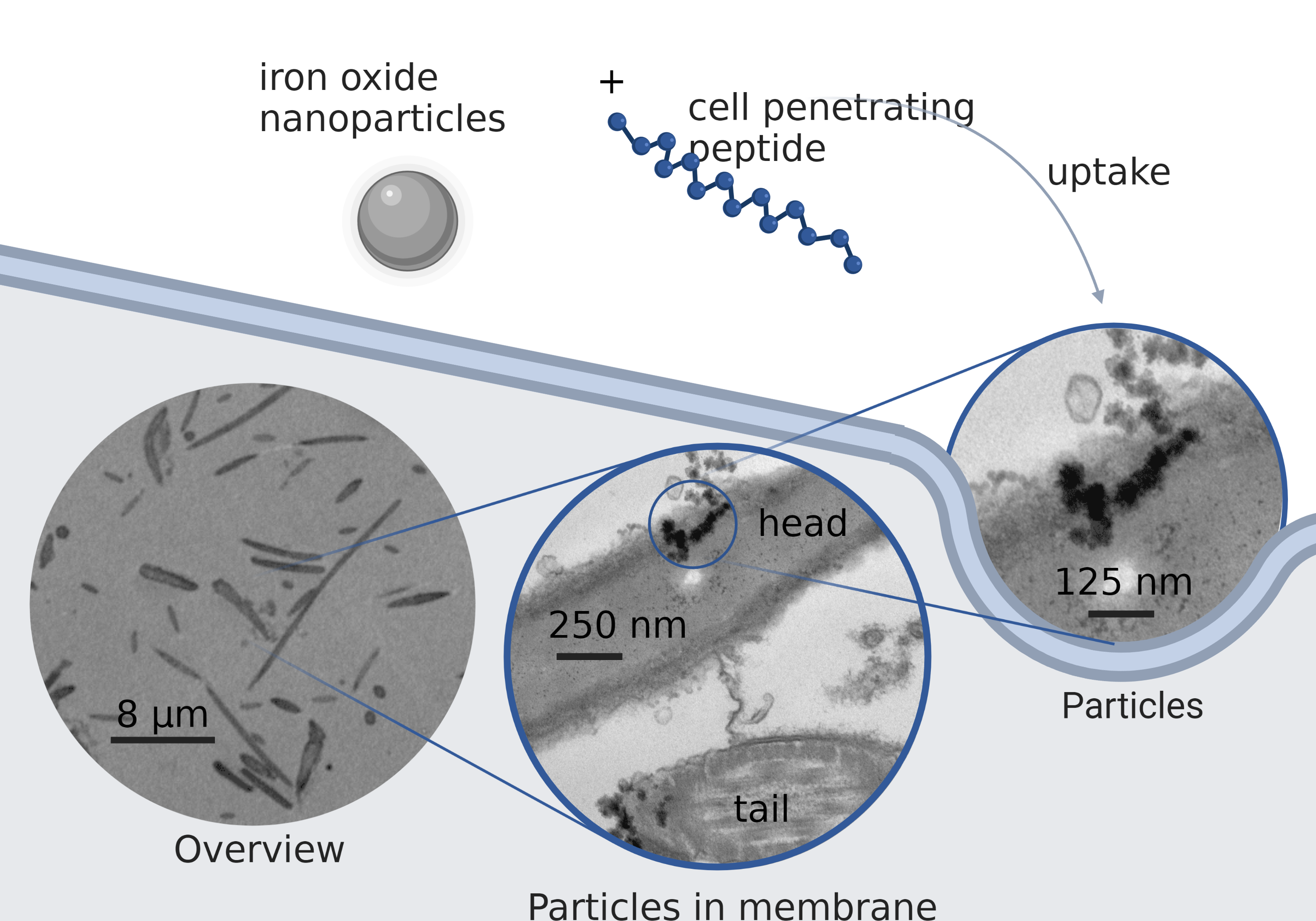}
\caption{\textbf{TEM images of cells stained with nanoparticles.} The figure illustrates how we believe the incubation worked. By incubating the sperm with iron oxide nanoparticles and Penetratin, the iron oxide nanoparticles internalized into the cell membrane (blue line). On the left is an overview of the sample, showing that the sperm are relatively close together, which is a possible cause for our inability to segment the tails in the synchrotron volumes. In the centre circle, the particle placement is shown. The particles were found both in the heads and tails as labelled in the image. The placement was determined by the texture of the image. On the right, we show a close up of the particles, where we see that several particles group together, forming lumps.}
\label{fig:em_results}
\end{figure}
To promote uptake of the iron oxide nanoparticles, we incubated a density gradient purified sperm sample with a suspension of the synthesized nanoparticles. The nanoparticles were incubated with the purified sperm using Penetratin to aid the uptake of the particles; see Materials and Methods. Then, the sperm samples were fixed and imaged with transmission electron microscopy (TEM) to verify that the particles had entered the sperm. The TEM images verified that the electron-dense magnetite nanoparticles had entered the cells (Figure~\ref{fig:em_results}). We determined the location of the particles to be inside the heads, midpieces and tails of the cells by investigating the texture of the cells in the EM-images. Traces of particles outside the cells were also observed.

\subsection{Heads and midpieces can be segmented from the synchrotron data}
\begin{figure}
\centering
\subfloat[Typical duroc sperm]{\noindent\includegraphics[width=0.99\linewidth]{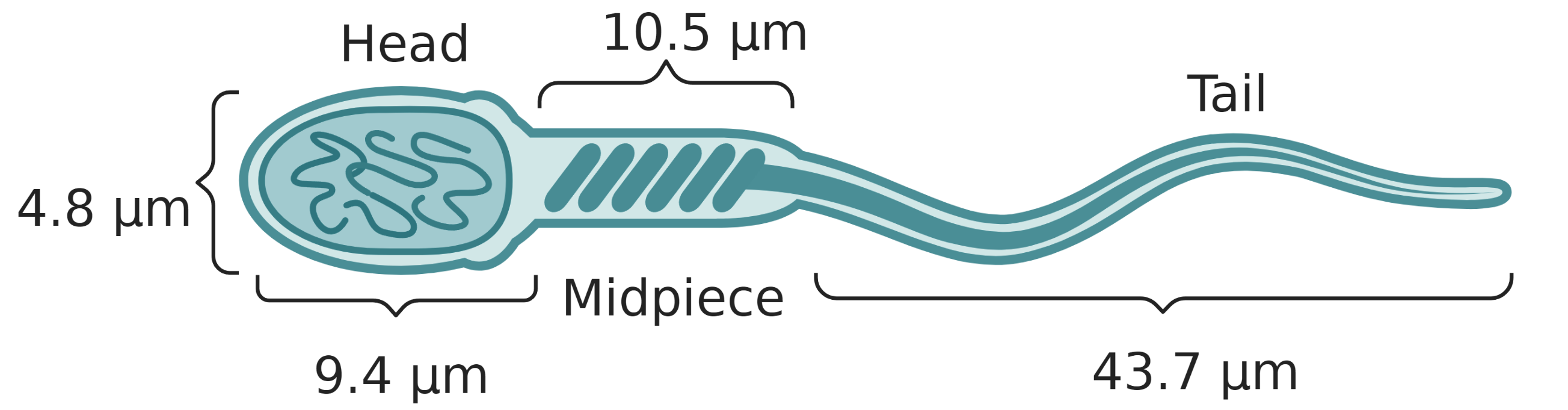}}\\
\subfloat[Synchrotron reconstructions and segmentations]{
\noindent\includegraphics[width=0.99\linewidth]{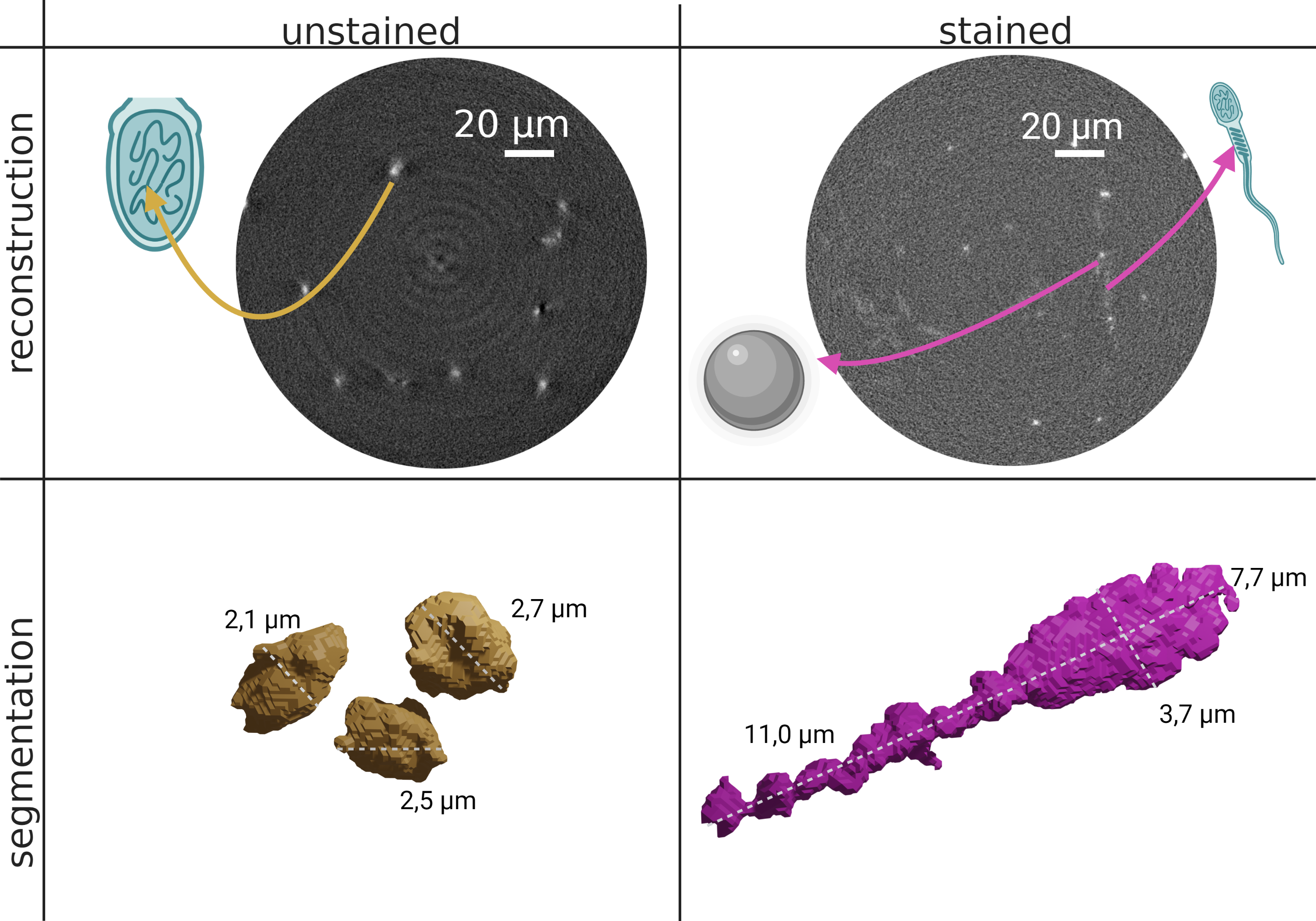}}
\caption{\textbf{Synchrotron results}. \textbf{A}: An illustration of a Duroc sperm with relevant mean measures. \textbf{B}: A tabular illustration showing the differences between unstained sperm (left column) and sperm stained with iron oxide nanoparticles (right column).  In the unstained sample (left column) we were not able to find any shapes reminiscent of sperm, only blobs we believe to be dense DNA inside the sperm heads. When the sperm were stained (right column) with our iron oxide nanoparticles the heads and midpieces could be segmented, which was verified by comparing the measures in the segmentation with the known morphological measurements of Duroc sperm (\textbf{A}). }
\label{fig:synchrotron_results}

\end{figure}

\begin{figure}
\centering

\subfloat[No CPP used]{
\noindent\includegraphics[width=0.49\linewidth]{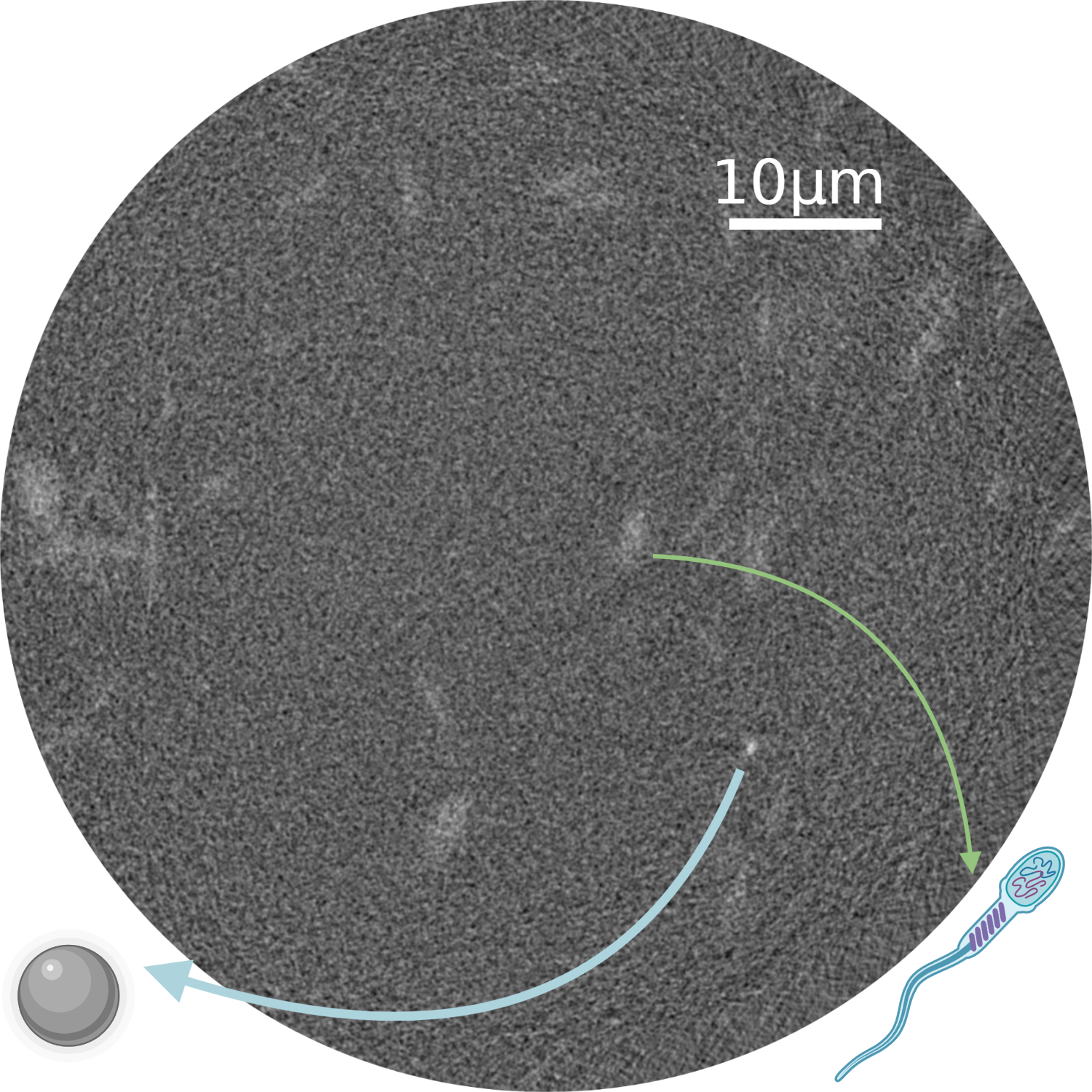}}
\subfloat[Penetratin]{
\noindent\includegraphics[width=0.49\linewidth]{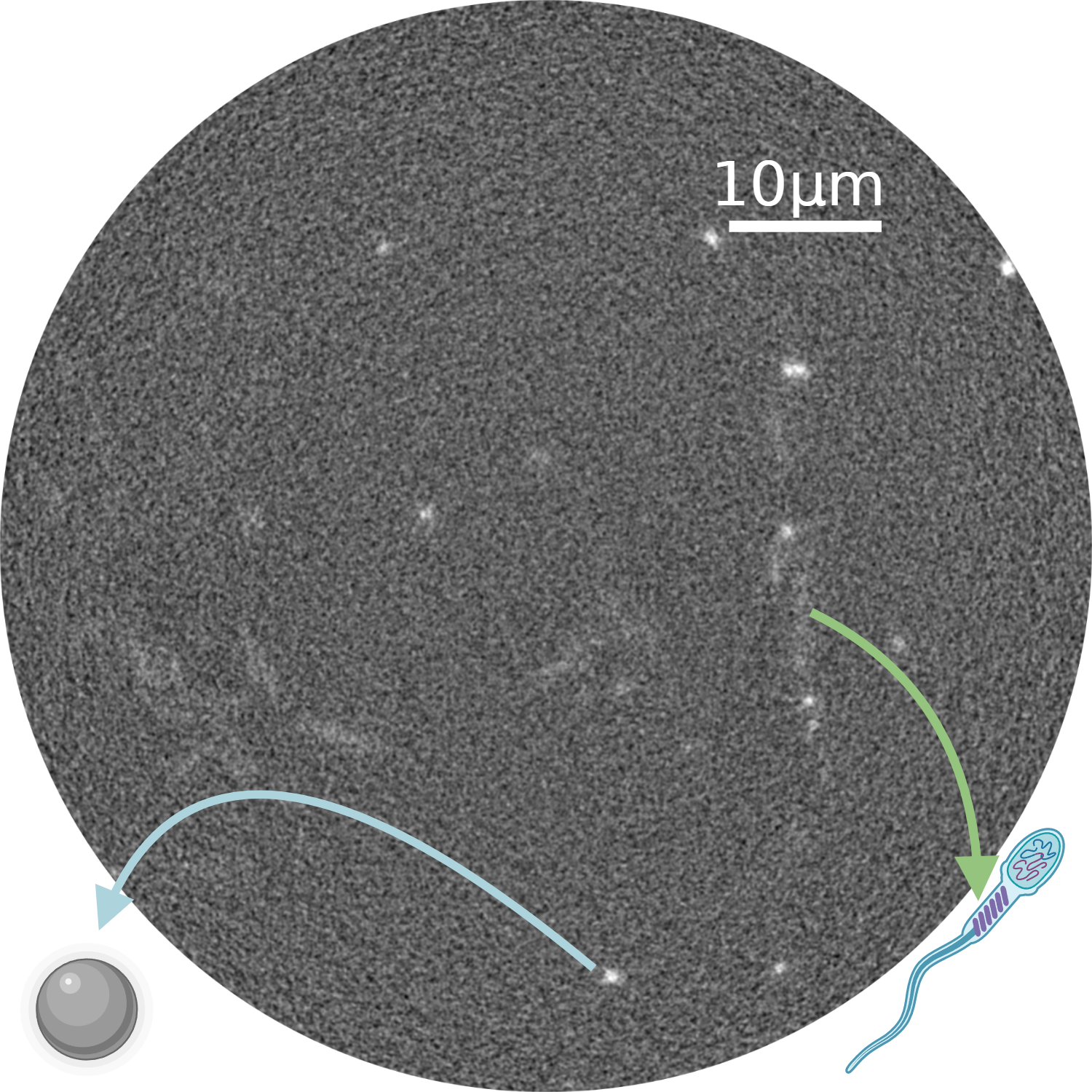}}
\caption{\textbf{Effect of penetratin.} \textbf{A}: The reconstruction slice from a sample stained with nanoparticles without the use of any CPP. \textbf{B} The reconstruction of a slice from a sample coincubated with iron oxide nanoparticles and Penetratin. In the sample where no CPP was used less high contrast areas could be found compared to the sample where the CPP Penetratin was used. Without the CPP we were only able to segment head like structures, while in the sample where the CPP Penetratin had been used we were able to segment cylindrical attachments to the heads consistent with the midpiece section of the cells.}
\label{fig:cpp_contrast}
\end{figure}

To assess the success of our contrast method, we performed SXRT of fixed sperm stained with nanoparticles and segmented the volumes.  We used morphological measurements of the sperm and compared them to our segmentations.\par
We used boar sperm of the Duroc breed to conduct our experiment as they are relatively robust for storage and transportation and readily available in large quatities.\par
Typical sperm, as illustrated in Figure~\ref{fig:synchrotron_results}(A), consists of a head, a midpiece, and a tail. The sperm of the Duroc boar that we used have been reported to have a flat oval head \SI{9.4\pm0.3}{\micro\metre} by \SI{4.8\pm0.2}{\micro\metre} and a tail \SI{43.7\pm1.2}{\micro\metre} long~\cite{Duroc}. To estimate the midpiece and tail morphology, we acquired a 3D TEM image stack of a Duroc sperm pellet and segmented it. From these segmentations, we estimate that the midpiece is about \SI{11\pm 0.1}{\micro\metre} long and \SI{0.9\pm 0.1}{\micro\metre} wide. We estimate that the tail is approximately \SI{500}{\nano\metre} in diameter at its widest place shortly after the neck and that the tail slowly decreases in width towards the tip to approximately \SI{300}{\nano\metre} in diameter. These measurements are consistent with the sperm morphology measures on domestic pigs in~\cite{Gu2019}.\par
We chose to conduct our experiments at the TOMCAT Nanoscope (SLS),
as it has a \SI{75}{\micro\metre} field of view, large enough to capture a whole cell and a \SI{65}{\nano\metre}
effective pixel size high enough to capture the tail. The samples we tested were all fixed Epon embedded samples as these are easier to work with than living hydrated samples. The density of water and Epon are both close to \SI[per-mode=symbol]{1}{\gram\per\milli\litre}, making the attenuation enhancement determined by the Epon samples comparable to what we would gain in a fluid sample. \par

We tested four samples at the synchrotron:
\begin{enumerate}
    \item Unstained cells.
    \item Cells that had been incubated with magnetite nanoparticles without the use of Penetratin.
    \item Cells that had been incubated with magnetite nanoparticles using Penetratin to aid the uptake.
\end{enumerate}

In sample 1, Figure~\ref{fig:synchrotron_results}(B), we were able to identify blob-like structures \SIrange[]{2}{3}{\micro\metre} in diameter, which we believe are dense DNA inside the sperm heads. We could not identify any structures resembling complete sperm heads or any tails structures in this sample. \par
In sample 2, we were able to identify disc like structures congruent with the flat oval shape of Duroc sperm heads, suggesting that the particles were taken up by the sperm and thereby increase contrast.\par
In sample 3, we were able to see the head and midpieces of the sperm. We detected several bright spots, consistent with enhanced contrast due to nanoparticles, surrounded by very low contrast lighter areas which we believe is the cell tissue. See Figure~\ref{fig:cpp_contrast} for a comparison of sample 2 and 3.\par

The segmentation of sample 3 showed oval disc-like structures, with a cylindrical attachment consistent with sperm heads and midpiece. The shape and size of the disc structures are consistent with the expected shape and size of a Duroc sperm head, and the length of the cylindrical attachment matches that of a Duroc sperm midpiece. The results are displayed in Figure~\ref{fig:synchrotron_results}(B).\par
The results from samples 2 and 3 suggest that magnetic nanoparticles can be used to enhance the contrast of sperm in synchrotron imaging at high energies. By taking advantage of the membrane permeating properties of the CPP Penetratin~\cite{helix} we were able to increase the uptake of magnetic nanoparticles and thereby the contrast.\par



\section{Discussion}
\emph{Achieving dynamic SXRT of sperm cells is a compromise between three factors; spatial resolution, temporal resolution and dose. Our results show that it is possible to use iron oxide nanoparticles for enhancing contrast in sperm cells during high-energy SXRT. Synchrotron imaging of live cells is extremely difficult as it poses a multitude of sample preparation difficulties, such as how to avoid bubbles in wet sample, how to catch a cell and store it in a small enough container to fit withing the field of view, how to protect it from radiation damage and how to obtain adequate contrast for analysis. While a non-toxic staining method does not solve all of these issues it is a necessary step towards imaging live cells in the future.}\par

Acquiring X-ray contrast at \SI{10}{\kilo\eV}, as we have done, requires labelling of the cells. Several works have shown that it is possible to increase contrast in X-ray projections by adding contrast enhancers such as osmium, gadolinium, iodine or gold nanoparticles~\cite{GadIod, Wichmann2013, nanoparticles, GoldNano, Cormode2011}. However, to the best of our knowledge, no other existing contrast method can be used with sperm cells without compromising their motility before the synchrotron scan. We also believe that the staining method could potentially be used to prepare other biological samples for SXRT, where survival of the specimen is important. The synthesis and internalization of iron oxide nanoparticles in sperm cells has previously been demonstrated,~\cite{Makhluf2006}, but the novelty of our work is demonstrating that by using the CPP Penetratin, it is possible to increase the contrast at high energy SXRT experiments enough to visualize sperm heads and midpieces.\par

Increasing the energy has the potential to decrease radiation damage to the sample. Two factors are important in lowering radiation damage; time and dose. The absorbed dose is measured as the absorbed energy pr. unit mass and is dependent on the X-ray mass attenuation coefficient of the sample. Higher mass attenuation is directly related to a higher dose~\cite{Attix2004}. The mass attenuation coefficients of soft tissue, similar in composition to sperm cells, decrease as energy increases~\cite{Seibert2005}. Hence, obtaining contrast of cells at higher energies is an essential step towards a viable tradeoff between dose and resolution for dynamic SXRT of live cells.\par

Since our future aim is to reconstruct the dynamic tail-beat of a sperm cell, an open question is whether the particles can also provide contrast in the tail region. As seen in the EM results in Figure~\ref{fig:em_results}, we found areas where several particles had entered the cells together instead of single particles evenly distributed. We observed particles in all regions of the cell, including the tail. \par

The synchrotron images corroborate the theory that the particles tend to group together inside the cells.The high contrast areas are larger in diameter than a single iron oxide nanoparticle (\SI{6}{\nano\metre}) and the resolution of the synchrotron (\SI{65}{\nano\metre}). This confirms that the particles tend to aggregate inside the cells. . The SXRT volume with stained cells also contains bright spots beside those attached to a head and midpiece, implying that some particles had either not entered the cells or were attached to the tails, which we could not segment. To determine whether the particles provide contrast in the tail region, we would ideally need to image a single cell with nanoparticles internalized devoid of extracellular particles. \par

The most significant limitation of the method presented in this study is the uneven distribution of iron oxide nanoparticles in the cells. Therefore, it is essential to determine how to get an even distribution of particles inside the cells to improve the method for visualization of sperm tail region. Also, the method is not meant as a tool for viewing the inner bio mechanics of the cells, but as a means to gain enough contrast to reconstruct the shape of the cell changing over time in 3D.\par

\emph{It has previously been shown that iron oxide nanoparticles have no adverse effect on sperm motility. In this article we show that iron oxide nanoparticles can be used as a contrast agent in SXRT imaging of sperm heads and midpieces. While the feasibility of our staining method is demonstrated on sperm, the method likely applies to many other cells as iron-oxide nanoparticles are widely used because of their low toxicity and biocompatibility. Our results in terms of spatial resolution also cover most other cells, as most animal and plant cells are \SIrange{1}{100}{\micro\metre} in diameter.}


\section{Methods and Material}

\subsection{Nanoparticle synthesis}
We synthesized iron oxide nanoparticles by modifying a previously reported co-precipitation method~\cite{babes}. Afterwards, a PVA solution was added to the nanoparticle suspension to incorporate a PVA surface coating~\cite{Makhluf2006}.\par

Iron oxide colloids were synthesized by dissolving \SI{2.08}{\gram} $\mathrm{FeCl_2 \cdot 4H_{2}O}$ (CAS: 13478-10-9) and \SI{0.80}{\gram} $\mathrm{FeCl_3}$ (CAS:7705-08-0) into a \SI{25}{\milli\litre} \SI{0.5}{\Molar} $\mathrm{HCl}$ (CAS: 7647-01-0) under an $\mathrm{N_2}$ atmosphere. Next, the acidic $\mathrm{Fe}$ salt solution was added dropwise to \SI{25}{\milli\litre} of an aqueous \SI{1.08}{\Molar} $\mathrm{NaOH}$ (CAS: 1310-73-2) solution under stirring. A PVA solution was made by dissolving \SI{2.52}{\gram} PVA (CAS: 9002-89-5) in \SI{50}{\milli\litre} deionized (DI) water under stirring for 30 minutes. The PVA solution was then added to the iron oxide suspension and stirred for another 30 minutes. The particles were washed 6 times with DI water, followed by precipitation by centrifugation at 4500 rpm  for 20 minutes, to remove $\mathrm{NaCl}$. Precipitated particles were re-dispersed in DI water and sonicated for 30 seconds in between washes.

\subsection{PXRD analysis}

We collected X-ray diffraction data from the iron oxide particles, the PVA solution, and a water background. Note that the iron oxide particles had not been thoroughly dried before measurements, and a scattering signal from water is therefore expected. The X-ray scattering pattern of magnetite was calculated with CrystalDiffract\textsuperscript{\tiny\textregistered}: a powder diffraction program for Mac and Windows. CrystalMaker Software Ltd, Oxford, England (www.crystalmaker.com). We used a transmission setup on an Empyrean Series 3 diffractometer from Panalytical with a Scintillation Detector and a $\mathrm{Ag K\alpha}$ X-ray source. Data from the particles and PVA solution were measured in 1.0 mm borosilicate capillaries, and data from water were measured in a 0.8 mm borosilicate capillary. The size of the magnetite crystallites was estimated with the Scherrer equation~\cite{scherrer, langford}, Equation~\ref{eq:YY}.

\begin{equation}
    L = \frac{K\cdot\lambda}{FWHM\cdot\cos(\theta)}\label{eq:YY}
\end{equation}

Where K is the Scherrer constant (0.9), $\lambda$ is the wavelength of X-rays used (\SI{0.56}{\angstrom}), FWHM is the width of the peak at half maximum ($2\theta=0.52^\circ$) and $\theta$ is the angle of the chosen peak ($2\theta=12.67^\circ$).\par

\subsection{Incubation of sperm with nanoparticles}
\begin{figure}
\centering
\includegraphics[width=\linewidth]{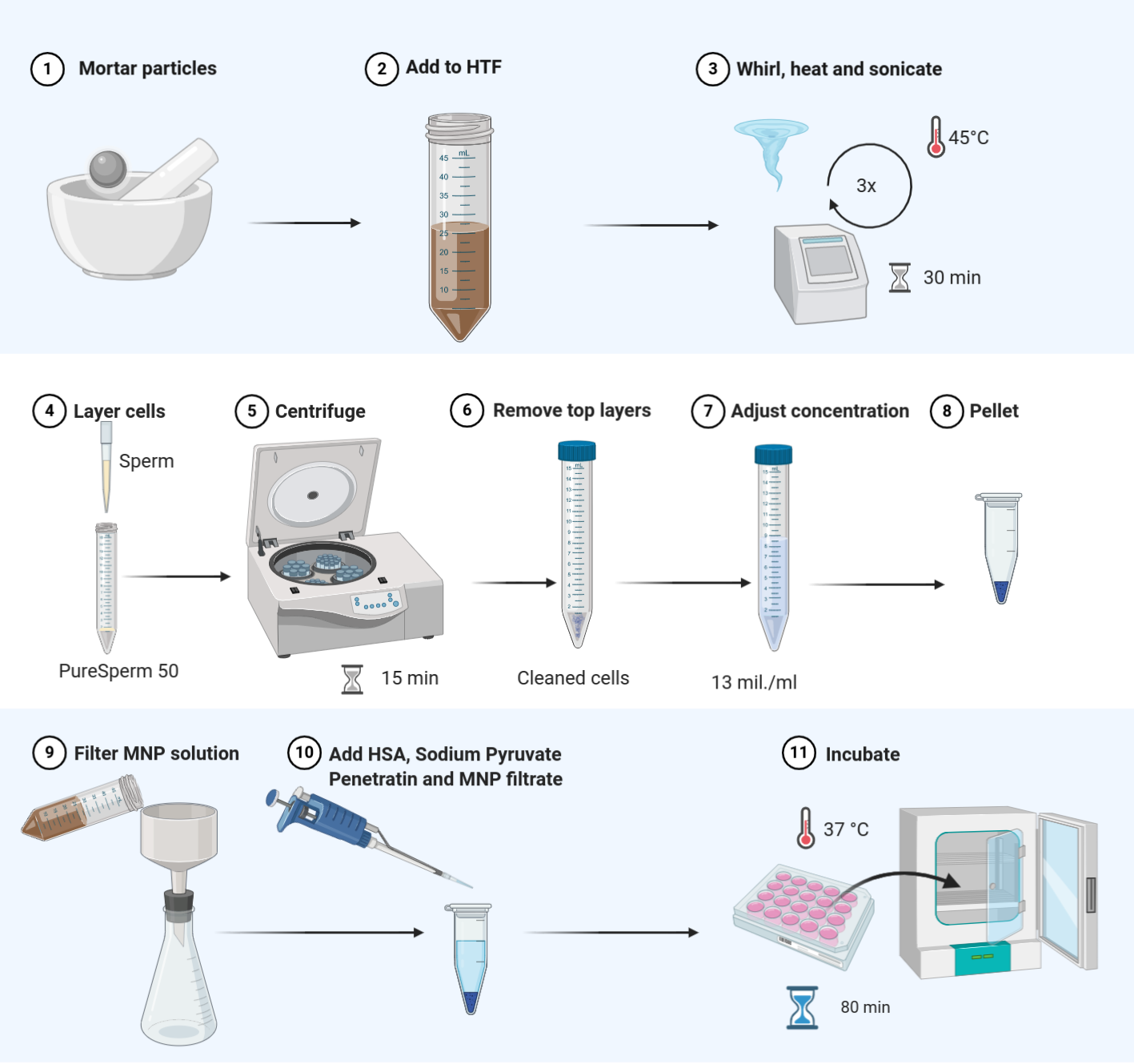}
\caption{\textbf{Overview of the incubation procedure.} The first row shows how to disperse the nanoparticles thouroughly in the HTF buffer suitable for sperm. The second row shows how to clean the sperm to get rid of objects otherwise attracting the iron oxide nanoparticles. The last row shows how to prepare the final incubation fluid and incubate the cells.}
\label{fig:attach}
\end{figure}
The rinsed magnetic nanoparticles were ground with a mortar to a fine powder and \SI{52}{\milli\gram} was added to \SI{20}{\milli\litre} \SI{4}{\milli\Molar} human tubular fluid (HTF) in a \SI{50}{\milli\litre} falcon tube. The suspension was then whirled for a short time at maximum power, heated to approximately \SI{45}{\degreeCelsius} and sonicated for 30 minutes. This was repeated three times until the particles were nicely dispersed in the liquid.\par
For our test, we used boar sperm of the Duroc breed. The sperm was cleaned prior to incubation using PureSperm 50\footnote{PureSperm 50: Mix 3 parts PureSperm 40 (Nidacon, PS40-100) and 1 part PureSperm 80 (Nidacon, PS80-100). E.g \SI{9}{\milli\litre} PureSperm 40 and \SI{3}{\milli\litre} PureSperm 80. Vortex. Store in refrigirator} to get rid of non-sperm objects which seem to attract the nanoparticles. First, \SI{1}{\milli\litre} of the sample was placed on \SI{2}{\milli\litre} cold PureSperm 50 in a \SI{15}{\milli\litre} falcon tube. Next, we spun the sample at \SI{400}{g} for 15 minutes at the lowest acceleration and deceleration and carefully removed the top 2 layers and $3/4$ of layer three such that approximately \SI{0.5}{\milli\litre} of the cloudy fluid remained above the pellet. We made HTF+ by adding \SI{150}{\micro\litre} sodium pyruvate to \SI{50}{\milli\litre} HTF buffer. The sperm sample was resuspended and diluted with HTF+ until the concentration was \SI{13e6}{\milli\litre\tothe{-1}}. We then made a pellet from \SI{1}{\milli\litre} of the rinsed sperm suspension by spinning 10 minutes at 700 rcf and removing the fluid. \par
Right before incubation, the nanoparticle suspension was filtered in a \SI{0.2}{\micro\metre}  filter to get rid of any large aggregates. After filtration we added \SI{15}{\micro\litre} sodium pyruvate per \SI{5}{\milli\litre} of the filtrate and \SI{30}{\micro\litre} Human Serum Albumin (HSA) per \SI{1}{\milli\litre} of the filtrate and vortexed the suspension. Finally, \SI{1.5}{\milli\litre} of the buffer solution containing nanoparticles was added to the sperm pellet and \SI{27}{\micro\litre} Penetratin stock for a final Penetratin concentration of \SI{5}{\micro\Molar}.\par
The sperm were incubated at \SI{37}{\degreeCelsius} for 80 minutes under foil and gentle. The final pellet was created by spinning at 700 rcf for 10 minutes, removing the top and slowly adding \SI{500}{\micro\litre} glutaraldehyde to the pellet for fixation. See Figure~\ref{fig:attach}. Due to the need to filter the nanoparticle suspension, it is not possible to accurately report the final nanoparticle concentration.\par
\FloatBarrier
\subsubsection{EM imaging}
Samples were fixed with $2\%$ v/v glutaraldehyde in \SI{0.05}{\Molar} sodium phosphate buffer (pH 7.2). Following centrifugation, the sample pellets were resuspended and rinsed in \SI{0.15}{\Molar} sodium cacodylate buffer (pH 7.2) three times. Next, the sample pellets were embedded in low-melting-point Agarose and postfixed in $1\%$ w/v $\mathrm{OsO_4}$ in \SI{0.12}{\Molar} sodium cacodylate buffer (pH 7.2) for \SI{2}{\hour}. The specimens were dehydrated in graded series of ethanol, transferred to propylene oxide and embedded in Epon according to standard procedures. Sections, approximately \SI{60}{\nano\metre} thick, were cut with an Ultracut UCT microtome (Leica, Wienna, Austria) and collected on copper grids with Formvar supporting membranes, stained with uranyl acetate and lead citrate, and subsequently examined with a Philips CM 100 TEM (Philips, Eindhoven, The Netherlands), operated at an accelerating voltage of \SI{80}{\kilo\volt} and equipped with an OSIS Veleta digital slow-scan 2k x 2k CCD camera (Olympus, Germany). Digital images were recorded with the ITEM software package.\par

\subsubsection{Synchrotron imaging}
The samples were prepared for the Nanoscope experiment by encasing them in EPON blocks, similar to the EM preparation but without any osmium staining. The blocks were then cut into small rods with a Dremel saw, such that the sample was centred in the rod. The rod was then inserted into the Dremel and rotated at minimum or close to minimum speed while carefully sanding the sample using sandpaper with decreasing grain sizes until the sample was below \SI{500}{\micro\metre} in diameter. These samples were imaged using SXRT at the TOMCAT nanoscope at \SI{10}{\kilo\eV}.\par

\emph{The TOMCAT Nanoscope is a high-resolution full-field transmission X-ray microscope based on Fresnel Zone Plate (FZP) focusing. This setup is working in the hard X-ray regime with energies from \SI{8}{\kilo\eV} to \SI{20}{\kilo\eV}. The instrument is composed of different optical elements: a custom designed beamshaper~\cite{Jefimovs2008}, an order selected aperture (OSA), a fresnel zone plate (FZP) and corresponding Zernike phase rings (ZPR). The beamshaper having a total diameter of \SI{2975}{\micro\metre} is made of iridium. It has an outermost zone width of \SI{60}{\nano\metre}, a square sub-field size with side of \SI{75}{\micro\metre} and radial periodicity of \SI{75.75}{\micro\metre} between each ring of sub-fields. It produces a top-flat illumination in its focal plane at 1684mm, at a distance of \SI{18.85}{\metre} from the X-ray source at \SI{10}{\kilo\eV}. The sample is then placed in his focal spot, having the size of the beamshaper sub-field (i.e. $75\times75\mu m^2$). The OSA is adjusted to keep only the first order of diffraction: the direct beam is stopped by a circular tungsten beamstop with a diameter of \SI{2}{\milli\metre}. The Fresnel Zone Plate (FZP) acts as an objective lens, i.e. magnifying the image on the detector. At \SI{10}{\kilo\eV}, we choose a zone plate in iridium with a diameter of \SI{204.6}{\micro\metre} and a focal length of \SI{99}{\milli\metre}. In order to enhance the contrast using Zernike phase contrast, Zernike phase rings are placed at the back-focal plane of the FZP. This is the usual contrast mechanism used for hard X-ray imaging with TXM to image low-Z materials such as biological tissue that usually are transparent to hard X-rays. Projections of the sample are collected onto the detector. In our case, the detector, consists of a Gadox scintillator screen (to convert X-rays to visible light), optical coupling fibres, and an sCMOS sensor (customized PCO
Edge 4.2 by Rigaku, \SI{6.5}{\micro\metre} pixel size) situated about \SI{10}{\metre} after the sample. The X-ray magnification in such a setup is obtained by the geometrical magnification, i.e. the ratio of the distance between the objective and the detector (in this case \SI{10}{\metre}) and the distance between the sample and the objective (in this case \SI{100}{\milli\metre}). With the TOMCAT nanoscope a constant magnification of x100 is obtained in the energy range \SIrange{8}{20}{\kilo\eV}, hence leading to an effective pixel size of \SI{65}{\nano\metre}.}\par

For the synchrotron samples we used an energy of \SI{10}{\kilo\eV} and collected 1500 projections over 180 degrees, associated with two series of 50 flats and one series of 20 darks. Exposure time was set to \SI{150}{\milli\second}. All reconstructions were achieved using the Gridrec reconstruction algorithm as implemented at the beamline~\cite{marone}. 3D datasets generated were saved in 16bits.

\subsubsection{Segmentation}
We segmented the cells in the reconstructed volume using scikit-image~\cite{scikit}. We used a standard Gaussian filter with a kernel size of 3.0 on all axis in the volume.

The volume was then thresholded using a triangle threshold~\cite{Zack1977}. Finally, we found all the connected components and deleted those containing less than 80000 voxels. The threshold of 80000 voxels was determined by visual inspection.\par

The visual illustrations in the article were created in BioRender.com.\par

\section{Author Contributions}

Conceptualization, M.B.L, A.B, T.D, K.A, J.S and M.K; Methodology, S.R.C, A.S.A, K.A, K.Q, and M.B.L; Formal Analysis, M.B.L, S.R.C, and A.S.A; Investigation, M.B.L, S.R.C, A.S.A, K.Q and A.B; Writing - Original Draft, M.B.L, S.R.C, A.S.A, K.Q, A.B, and M.K; Writing - Review \& Editing, A.B, T.D, K.A, J.S, M.K, A.S.A, S.R.C, M.B.L, K.Q and K.M.Ø.J; Visualization, M.B.L, A.S.A and S.R.C; Supervision, J.S, T.D and K.M.Ø.J; Funding Acquisition, J.S., T.D. and K.M.Ø.J. 
\section{Acknowledgement}
MBL was supported by the Capital Region Research Foundation (Grant number: A5657) on the MAX4Imagers project (PI: Tim Dyrby).\par
ASA, SRC and KM\O J are grateful to the Villum Foundation (grant number: VKR00015416) and the Carlsberg Foundation (grant number: CF17-0976) for funding.\par
We acknowledge the Paul Scherrer Institut, Villigen, Switzerland, for the provision of synchrotron radiation beamtime (20201724) at the beamline TOMCAT of the Swiss Light Source.




\printbibliography



\end{document}